\documentstyle[aps,preprint]{revtex}
\begin{document}
\draft
\title{Continuum Model for River Networks}
\author{Achille Giacometti$^{(1)}$, Amos Maritan$^{(2)}$,
and Jayanth R. Banavar$^{(3)}$}
\address{$^{(1)}$ Institut f\"{u}r Festk\"{o}rperforschung
des Forschungszentrums J\"{u}lich, \\
Postfach 1913, D-52425, J\"{u}lich, Germany}
\address{$^{(2)}$International School for Advanced Studies \\
and Istituto Nazionale di Fisica Nucleare INFN , \\
via Beirut 2-4, 34014 Trieste, Italy}
\address{$^{(3)}$Department of Physics and Center for Materials Physics,
104 Davey Laboratory,\\
The Pennsylvania State University, University Park, PA 16802}

\date{\today}
\maketitle
%\addtolength{\baselineskip}{\baselineskip}
%%%%%%%%%%%%%%%%%%%%%%%%%%%%% Abstract %%%%%%%%%%%%%%%%%%%%%%%%%%%%%%%%%%
\begin{abstract}
The effects of erosion, avalanching and random precipitation are
captured in a simple stochastic partial differential equation for
modelling the evolution of river networks.  Our model leads to a
self-organized structured landscape and to abstraction and piracy of the
smaller tributaries as the evolution proceeds.  An algebraic
distribution of the average basin areas and a power law relationship
between the drainage basin area and the river length are found.
\end{abstract}
%%%%%%%%%%%%%%%%%%%%%%%%%%%%% Text %%%%%%%%%%%%%%%%%%%%%%%%%%%%%%%%%%%
\newpage
\narrowtext
A fractal river network is a striking example of self-organized
criticality.  The physics of river network evolution arises from an
interplay of the structured landscape governing the water flow with the
erosional effects of the water feeding back into further sculpting of
the landscape. Extensive studies of the fractal characteristics of real river
networks have been carried out. \cite{rrrb,hack,lm,sb,glock,scheid,horton}
Hack\cite{hack} has studied the relationship
between the length of a river $l$ and the area of a drainage basin $s$.
$s$ is a measure of the total area of the land covered by the principal
stream and its tributaries that feed into the network.  Hack's
measurements indicate that for basin areas $s$ ranging over almost five
decades (up to 375 square miles), $s \sim l ^{\phi}$ with the exponent
$1/\phi \sim 0.57$ .  Other measurements of the distribution
of drainage basin areas suggest a power law scaling of the form $P(s)
\sim s^{-\tau}$ with
$\tau=1.45 \pm 0.03$ . \cite{rrrb}

Most of the models of river networks fall into two categories.  The first is
restricted to reproducing the statistical properties of
networks \cite{gm}.  More
recently, models for the {\it evolution} of river networks have
been developed.  Based
on careful studies of river data \cite{rrrb}, Rinaldo and
coworkers\cite{rrri} have suggested
that the effects of local optimal rules equivalent to critical erosion
parameters lead to statistical characteristics for the networks similar
to global constraints of minimum energy dissipation.
Leheny and Nagel \cite{ln} introduced a lattice model that
incorporated erosion and showed a competition in growth between
neighboring river basins relevant for late stages of evolution.
Kramer and Marder\cite{km}
have also constructed a lattice model that allows for the elucidation of
the scaling properties of the large scale features of river networks.
In addition, they have proposed coupled differential equations for two
scalar fields, the height of the soil and the depth of the water flowing
over the soil.  An analysis of these equations has led to an
understanding of the shape and stability of individual river channels.

Our principal goal is to introduce and numerically study a simple
stochastic partial differential equation for the evolution of the
landscape of a river network.  Our model is a field theory for the soil
height $h(\vec{x}, t)$ and takes into account the effects of random
precipitation, erosion and the avalanching of soil.  An initially smooth
landscape evolves into a nontrivial spatially self-organized state in
which Hack's law and the algebraic distribution of drainage basin areas
are obtained.

The evolution equation may be written in the compact
form:
\begin{equation} \label{evolution}
\frac{\partial}{\partial t} h(\vec{x} ,t) = {\cal D} \cdot h(\vec{x}
,t) - k \nabla ^{4} h(\vec{x} ,t) + \eta (\vec{x} ,t)
\end{equation}

\noindent
where ${\cal D} \cdot  \equiv D _{1} \partial^{2}_{y} + \tilde{D}_{2}
(\mid \nabla h \mid ) \partial ^{2}_{x}$ and $\vec{x} \equiv (x,y)$.  The
coefficient $\tilde{D}_{2} (\mid \nabla h \mid)$ is defined as:
\begin{equation}
\tilde{D}_{2} [\mid \nabla h \mid ] = \left\{ \begin{array}{ll}
D_{1} > 0 & \mbox{if} \mid \nabla h \mid > M \\
D_{2} < 0 & \mbox{if} \mid \nabla h \mid < M
\end{array}
\right.
\end{equation}

\noindent
and the random noise $\eta (\vec{x},t) \equiv \varepsilon r (\vec{x},t)$ is
Gaussianly distributed with zero average and correlation
\begin{equation}
< \eta (\vec{x},t) \eta (\vec{x}' ,t') > = \varepsilon ^{2} \delta ^{d}
(\vec{x}-\vec{x}')
\delta (t-t')
\end{equation}

The equation describes the temporal evolution of a two dimensional
landscape with periodic boundary conditions in the $x$ direction and
a dominant water flow in the $y$ direction due to an incline.
The equation
allows for ordinary diffusion in the $y$ direction and a stabilizing
$\nabla ^{4} h$ term working as an ultraviolet regulator.
In the $x$ direction, a diffusion term is
operational as long as $\mid \nabla h \mid > M$.  These diffusional
processes mimic the avalanching effect discussed by Leheny and Nagel:
when neighboring sites have large differences in heights, avalanching of
the soil occurs leading to a smoothing effect.  The erosional processes
are captured by the negative diffusion coefficient $D_{2}$ when $\mid
\nabla h \mid < M$.  Such a term accentuates the height difference and
captures the physics of erosion - the water flowing in the shallower
parts of the landscape erode the soil further, {\it increasing}
the height difference. In the simple version of our model, the erosional
processes are blind - the erosion is uncorrelated with the water flow.
Also, our analysis has been carried out in the limit of a large
underlying $y$ slope so that the water flow is directed downhill and
ordinary diffusion is operative in the $y$ direction.
Large basins are known to have smaller overall slopes so that a less
directed water path would be more appropriate.\cite{rrri}
The noise $\eta$ mimics the erosional effects of random precipitation.
It should be noted that this noise is an essential ingredient of
the dynamics: noisy initial conditions would evolve into a flat landscape
under only the deterministic part of equation (\ref{evolution}).

Our equation is a generalization of equations developed in other
contexts.  In the limit that $D_{1} = D_{2} > 0$, the equation reduces
to the Edwards-Wilkinson equation for linear growth processes \cite{ew}.
When $D_{1}$ and $D_{2}$ are both positive but different, the equation is akin
to Barenblatt's equation \cite{baren} describing
the pressure in an elasto-plastic
porous medium that allows for the contraction and expansion of the
medium due to the flow.  The novel ingredient in our equation is the
possibility of a negative diffusion coefficient.  As in sandpile models
of self-organized criticality \cite{bak}, smoothing
diffusional processes kick in
when the gradient exceeds a critical value.  While the sandpile models
are driven by the random {\it addition} of sand, here the
instability is caused by the erosion and the noise.  Our model evolves in
a self-organized manner
to a noisy boundary separating stability and instability, as do the
sandpile models.   In the sandpile case, there is a well-defined interval
between the addition of grains during which the relaxation takes place.
In contrast, the dynamics is continuously turned on in our model.

In order to discretize eq. (\ref{evolution}) it is convenient to
work with dimensionless variables $\tilde{t},\tilde{\vec{x}},\tilde{h}$
defined as:
\begin{equation}
t = \frac{k}{D^{2}_{1}} \tilde{t}
\end{equation}
\begin{equation}
\vec{x} = {\sqrt \frac{k}{D_{1}}} \tilde{\vec{x}}
\end{equation}
\begin{equation}
h = \frac{\varepsilon}{\Delta x {\sqrt D_{1}}} \tilde{h}
\end{equation}
where $\Delta x$ is the lattice spacing of the spatial mesh we will be
using. In the following we shall omit the tildas.
The discretized version of the equation is:
\begin{eqnarray}
h(\vec{x},t + \Delta t) &=& h(\vec{x},t) + \frac{\Delta t}{\Delta x^{2}}
(\Delta_y h(\vec{x},t) + D[\mid \nabla h \mid] \Delta_x h(\vec{x},t))
\\ \nonumber
 &-& \frac{\Delta t}{\Delta x^{4}}
\Delta^2 h(\vec{x},t) + \sqrt{\Delta t} \; r(\vec{x},t)
\end{eqnarray}

\noindent
Here $\Delta_x$ and $\Delta_y$ are discrete second derivatives in
the $x$ and $y$ direction and $\Delta^2$ is the discrete square laplacian
defined so that the Fourier transform of $\Delta^2 h$ does not
contain anisotropic terms to the leading order, and
\begin{equation}
D [\mid \nabla h \mid ] = \left \{ \begin{array}{ll}
1 & \mbox{if} \mid \nabla h \mid >M\\
D \equiv \frac{D_{2}}{D_{1}} < 0 & \mbox{if} \mid \nabla h \mid <M
\end{array} \right .
\end{equation}

Once the landscape reaches a self-organized state, we obtain a measure
of $s$ by adding a test drop of precipitation on each site and following
its downhill descent along the steepest path \cite{rrrb,ln}.  A drop at
site ($x$,$y$) has three possible destination sites - ($x-1$,$y+1$),
($x$,$y+1$) or ($x+1$, $y+1$).  The site with the smallest $h$ is selected.
On a given site, $s$ is defined as the number of drops collected on
that site during the downhill evolution.

If we define $P(s,L)$ as the density of sites with a total flow $s$
on a landscape of linear size $L$,
one may combine Hack's law and the algebraic scaling of the river basin
areas into a generalized scaling form \cite{maritan}:
\begin{equation} \label{scaling}
P(s,L) = s^{-\tau} F(s/L ^{\phi'})
\end{equation}
Assuming that rivers are fractals with length $l \sim L^{d_F}$, the
characteristic
basin area $s_c \sim L^{\phi'} \sim l^{\phi'/d_F}$ implying  Hack's
law with $\phi=\phi'/d_F$. In the present case we find $d_F=1$,
i.e. the rivers are self-affine, and thus $\phi=\phi'$.

{}From the definition of $s$, it follows that
its average $\sum_s P(s,L) s =(L+1)/2$ which combined with (\ref{scaling})
leads to the scaling law  $\phi = (2-\tau )^{-1}$. \cite{meakin}
Further, for $\tau >1$, the normalizability of
$P(s,L)$ in the $L \rightarrow \infty$ limit imposes a constraint that
$\lim_{x\rightarrow 0} F (x)$ is universal and equal to
$(\tau -1)$.  Figure \ref{fig1} shows a plot of $P(s,L)$ for different sizes
$L=$ 10,30,50 and 100 at time $t=$10. These results were obtained
in the regime where $\Delta t/\Delta x^2 \sim 10^{-3}$, well below
the linear stability limit $\Delta t/\Delta x^2 \sim 0.25$.
The figures are then collapsed
into a single scaling plot (Figure \ref{fig2}) with the choices of $\tau = 4/3$
and $\phi = (2-\tau) ^{-1} = 3/2$.\cite{note}
Within the error estimates of our
analysis, these exponents are the same as obtained in Scheidegger's
static model\cite{scheid} for river networks \cite{nagatani}.
They are also within the range
observed for real rivers in the small basin limit.
It should be mentioned that the above scaling form is expected
to be valid only in the large $s$ and large $L$ regime.
Long time runs on small systems are suggestive of a non-Scheidegger
universality class at large times.  However, a quantitative study of this
regime
appears to be beyond our present capability due to computational limitations.

We have also monitored the temporal evolution of the roughness of the
landscape. The roughness $W$ is defined as:
\begin{displaymath}
W(t) = ( \frac{1}{L^2} \sum_{\vec{x}}
< (h(\vec{x},t)-\overline{h}(t))^2>)^{1/2}
\end{displaymath}
where $\overline{h}(t) =\sum_{\vec{x}}h(\vec{x},t)/L$. The average was
performed over 300 samples corresponding to different realizations
of the noise.
We find an intermediate regime with an algebraic growth $W(t) \sim t^{\beta}$
with an exponent $\beta = 0.21 \pm 0.02$ (Figure \ref{fig3}).
This exponent value is different from the exactly
solvable Edwards-Wilkinson model \cite{ew} where $\beta=0$, i.e.
$W(t) \sim \sqrt{\ln t}$.
At longer times saturation due to finite size is observed.
In this context, it is interesting to note the recent work of
Czirok et. al. \cite{czirok} on a geomorphological micromodel of
mountains.
The evolution of river networks is shown in Figure \ref{fig4}, where only sites
with $s \geq 30$ are shown, while the corresponding elevation profile
is shown in Figure \ref{fig5}.  The model captures some of the key ingredients
of Glock's theory of the evolution of river networks\cite{glock} -
small tributaries
are eliminated as the main rivers swell in size.  The figures also show
the effects of piracy in which a larger aggressive stream captures the flow
of a neighboring stream.  Such effects had also been shown to occur by Leheny
and Nagel\cite{ln} in their lattice model for river networks.

A more realistic model than the one presented here would have coupled
differential equations for the variables
$h(\vec{x},t)$ and the flow $s$. Such a coupling would capture
non-local effects present in real basins.

We are indebted to Andrea Rinaldo and Joachim Krug for enlightening discussions
and the former for a critical reading of the manuscript.
This work was supported by grants from INFN, NASA, NATO, NSF, ONR, EPSRC,
The Fulbright Foundation,
The Petroleum Research Fund administered by the American
Chemical Society and the Center for Academic Computing at Penn State.
AG acknowledges partial support from the HCM program
under contract ERB4001GT932058.
%\newpage

%\begin{thebibliography}{99}

%\newpage
%\begin{center}
%\large
%{\bf Figure Captions}
%\end{center}
%\normalsize
%Fig1
\begin{figure}
\caption{Plot of $P(s)$ vs. $s$ for sizes $L=10,30,50,100$ evaluated at
the (absolute) time $t=10$ in $d=2+1$.  The values for the parameters
are: $\Delta t=0.01, \Delta x=3.0, M=1, D_{2}/D_{1} = -0.1$.}
\label{fig1}
\end{figure}
%Fig2
\begin{figure}
\caption{Collapse of the curves of $P(s)$ from Figure 1 with $\tau = 4/3,
\; \phi = (2-\tau)^{-1}=3/2$.}
\label{fig2}
\end{figure}
%Fig3
\begin{figure}
\caption{Temporal evolution of roughness $G_2(t)=W^2(t)$.
The parameter used
for the simulations were $\Delta t=10^{-4}$, $\Delta x=0.5$, $M=1$,
$D_{2}/D_{1} = -0.1$.}
\label{fig3}
\end{figure}
%Fig4
\begin{figure}
\caption{ A snapshot of a typical river network created by
our dynamics at two successive times $t$ (a) and $t'>t$ (b).
Only sites with $s \ge 30$ are shown. The values of $t$ and $t'$
correspond to $89980$ and $90000$ temporal iterations with
$\Delta t=10^{-4}$, $\Delta x=0.5$, $M=1$, $D_{2}/D_{1} = -0.1$.
The initial condition was a flat configuration. The flow
is constructed as follows: starting from a site $(x,y)$
the water can flow only to one of the three sites $(x+1,y+1)$,$(x,y+1)$
or $(x-1,y+1)$ thus causing an overall flow downward (from $y$ to
$y+1$). The choice of which of the three sites is determined
by the relative heights of the corresponding $h$'s
(steepest descent). In the region indicated by P one main stream
which was present in  (a) has been captured by a neighboring main
stream in (b) (piracy effect). An example of a stream present in (a)
which has disappeared in (b) (abstraction effect) is also
shown in the region A.
The boundary conditions  employed were periodic
in the $x$ direction (perpendicular to the flow) and free
in the $y$ direction (parallel to the flow).}
\label{fig4}
\end{figure}
%Fig5
\begin{figure}
\caption{Landscape (a) and contour plot (b) at $t'$ corresponding to Fig.4(b).
Values for the parameters are the same as Fig.4(b)}
\label{fig5}
\end{figure}

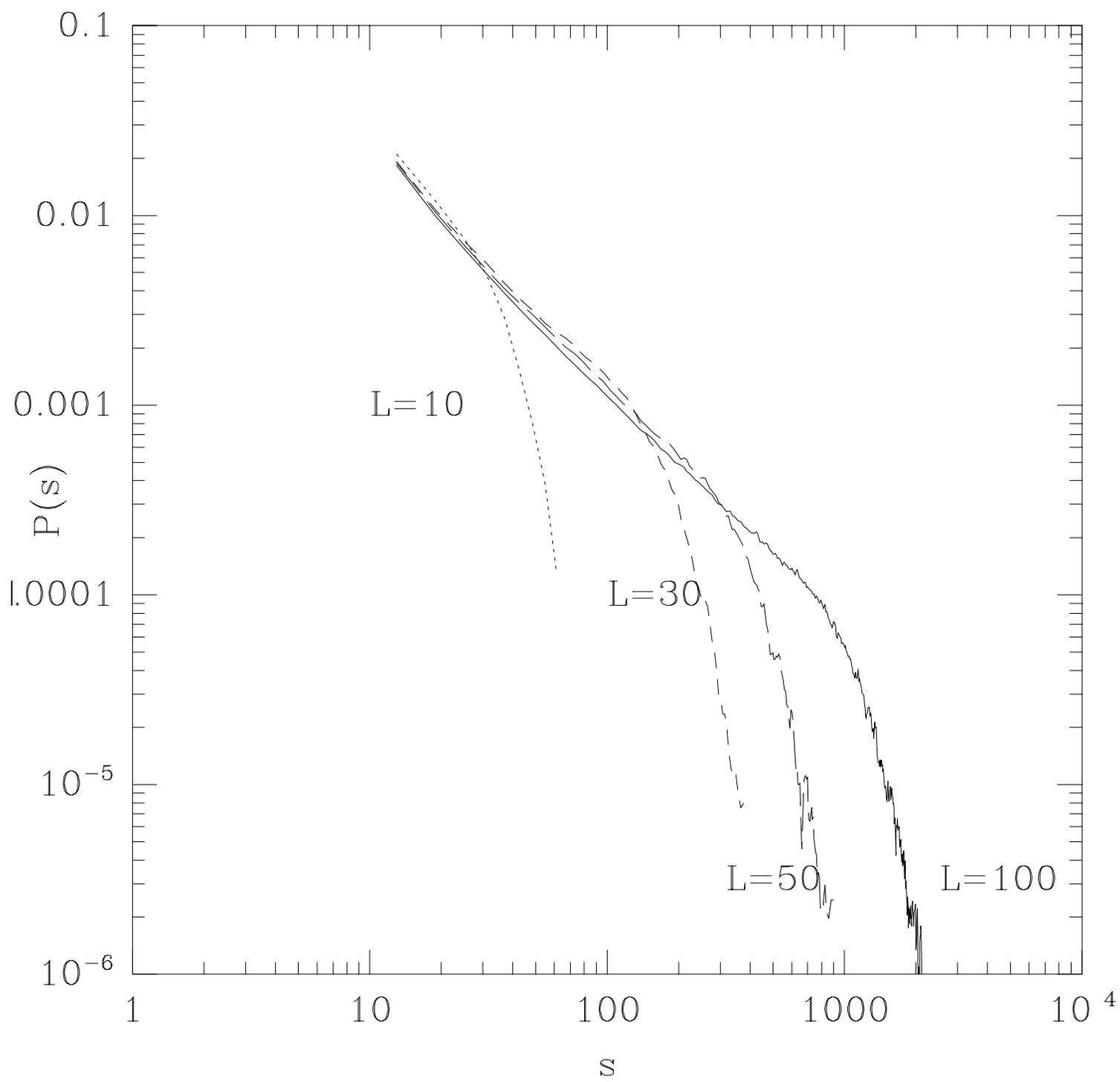
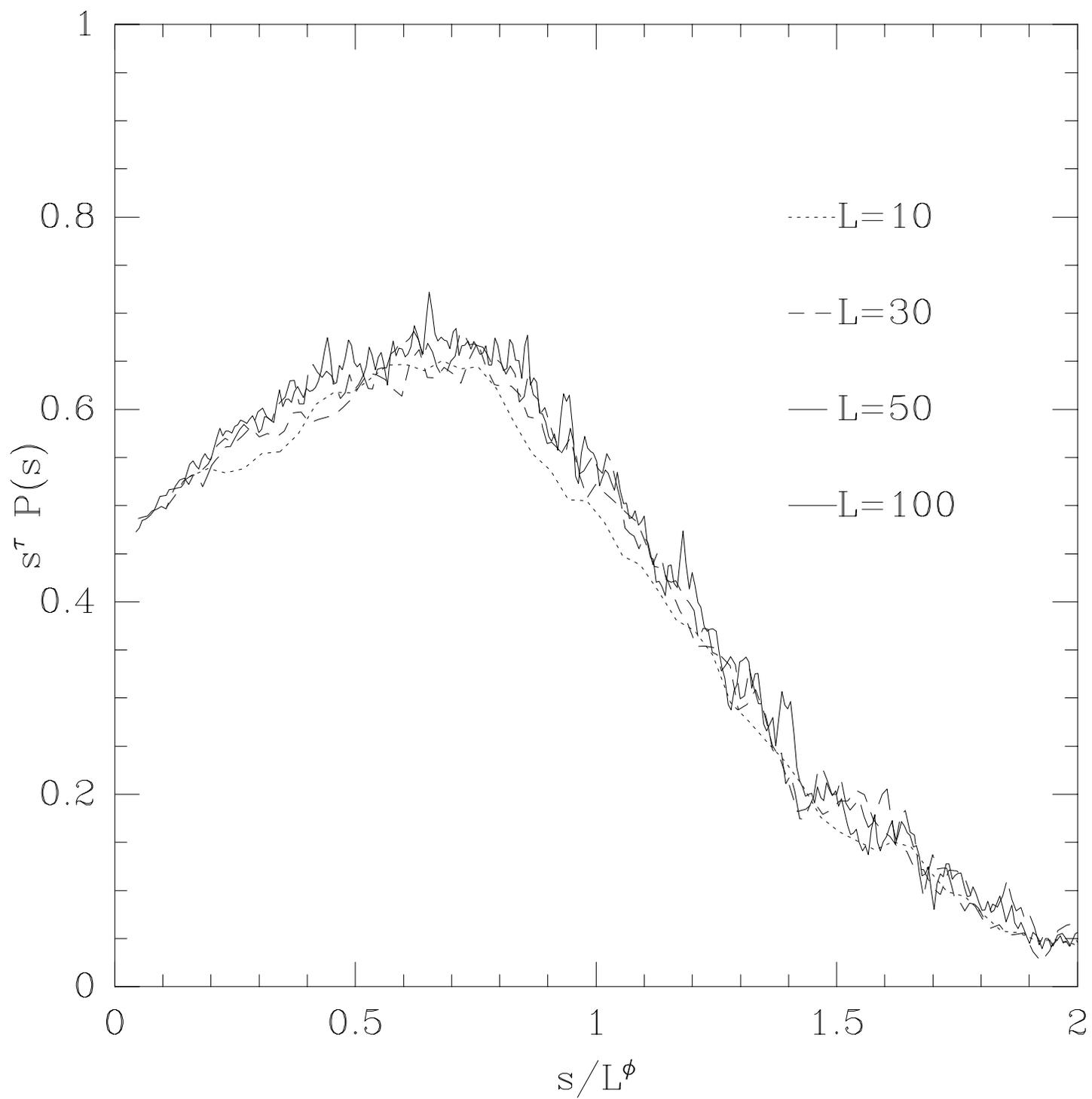
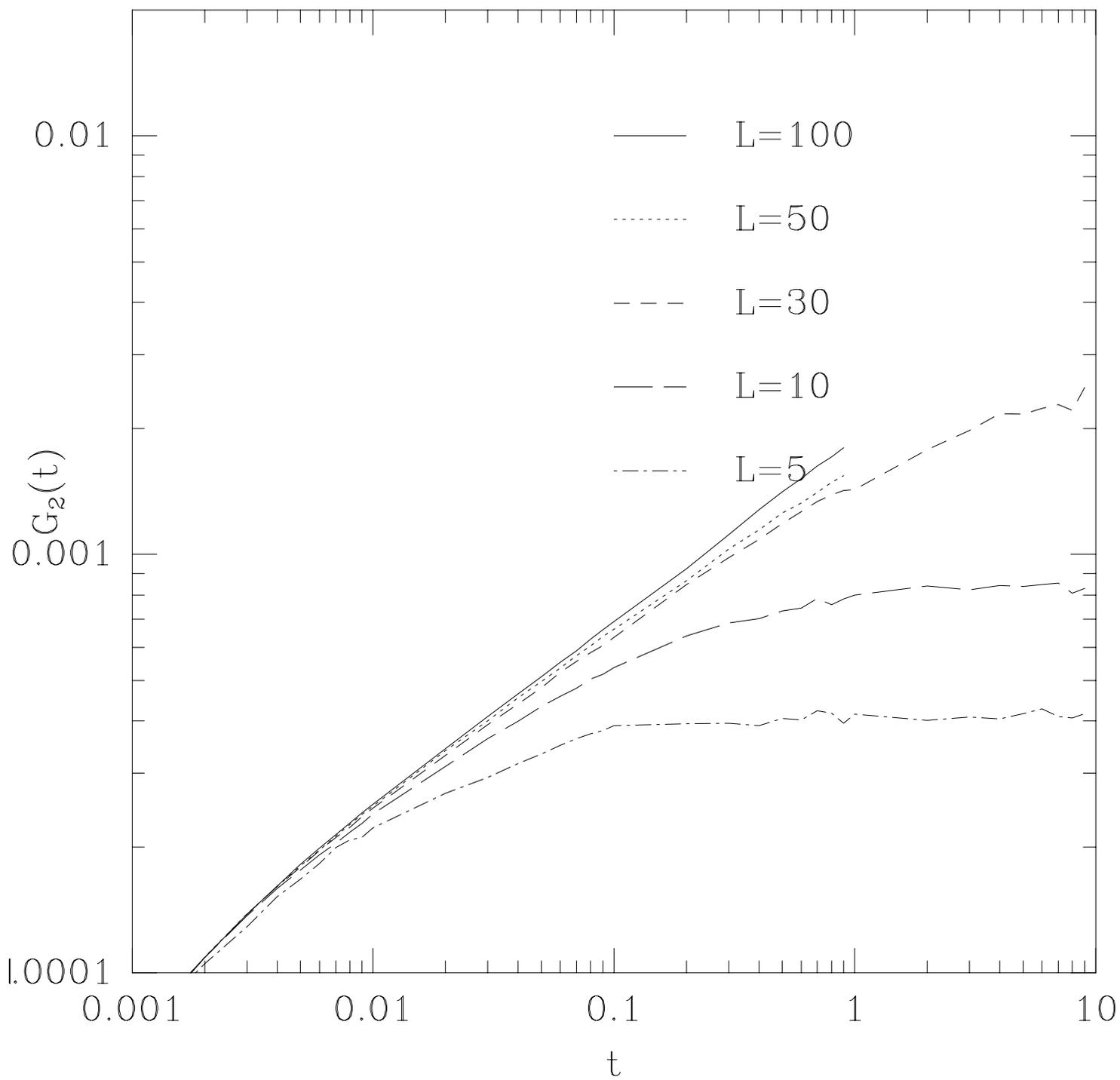
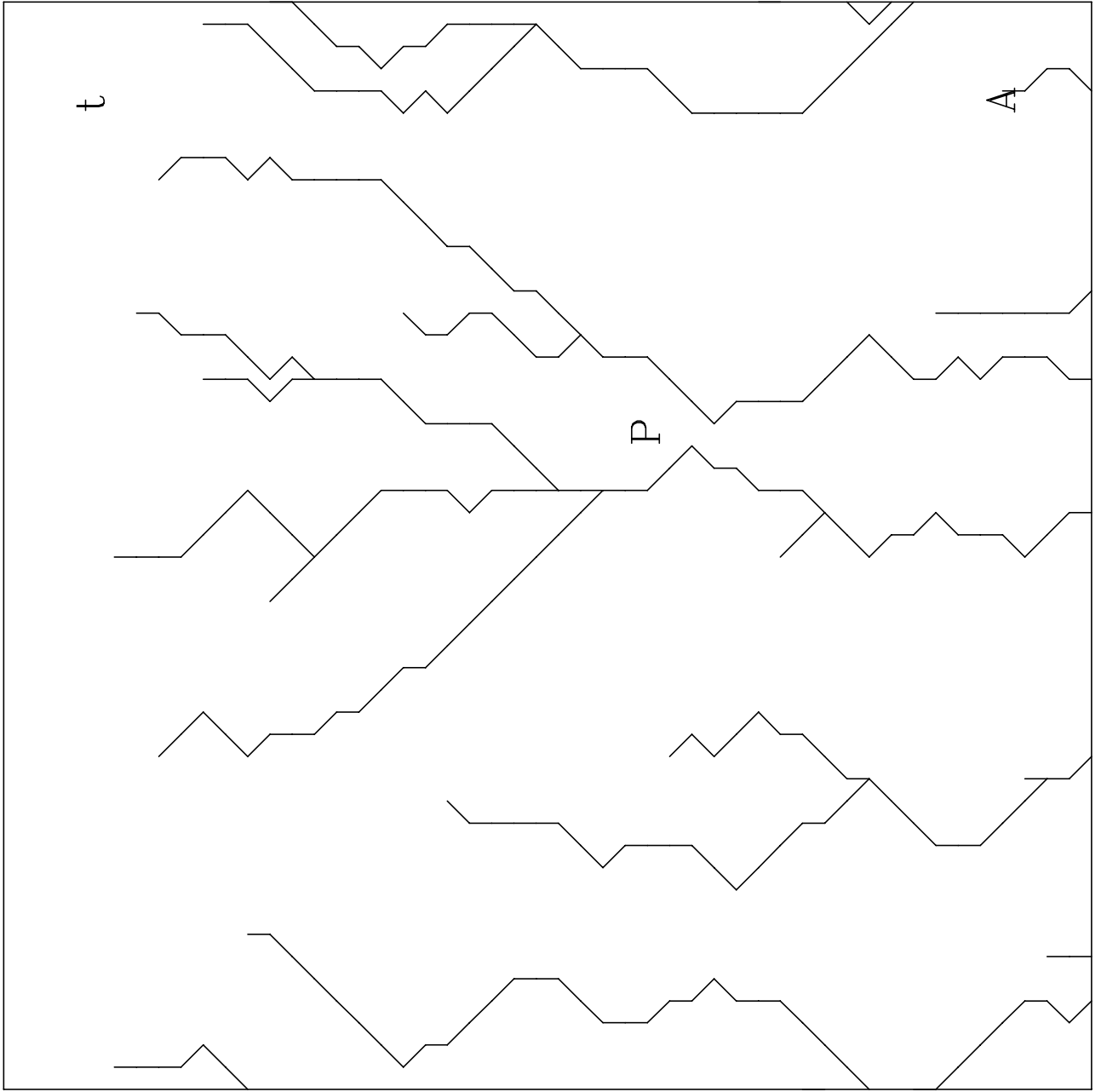
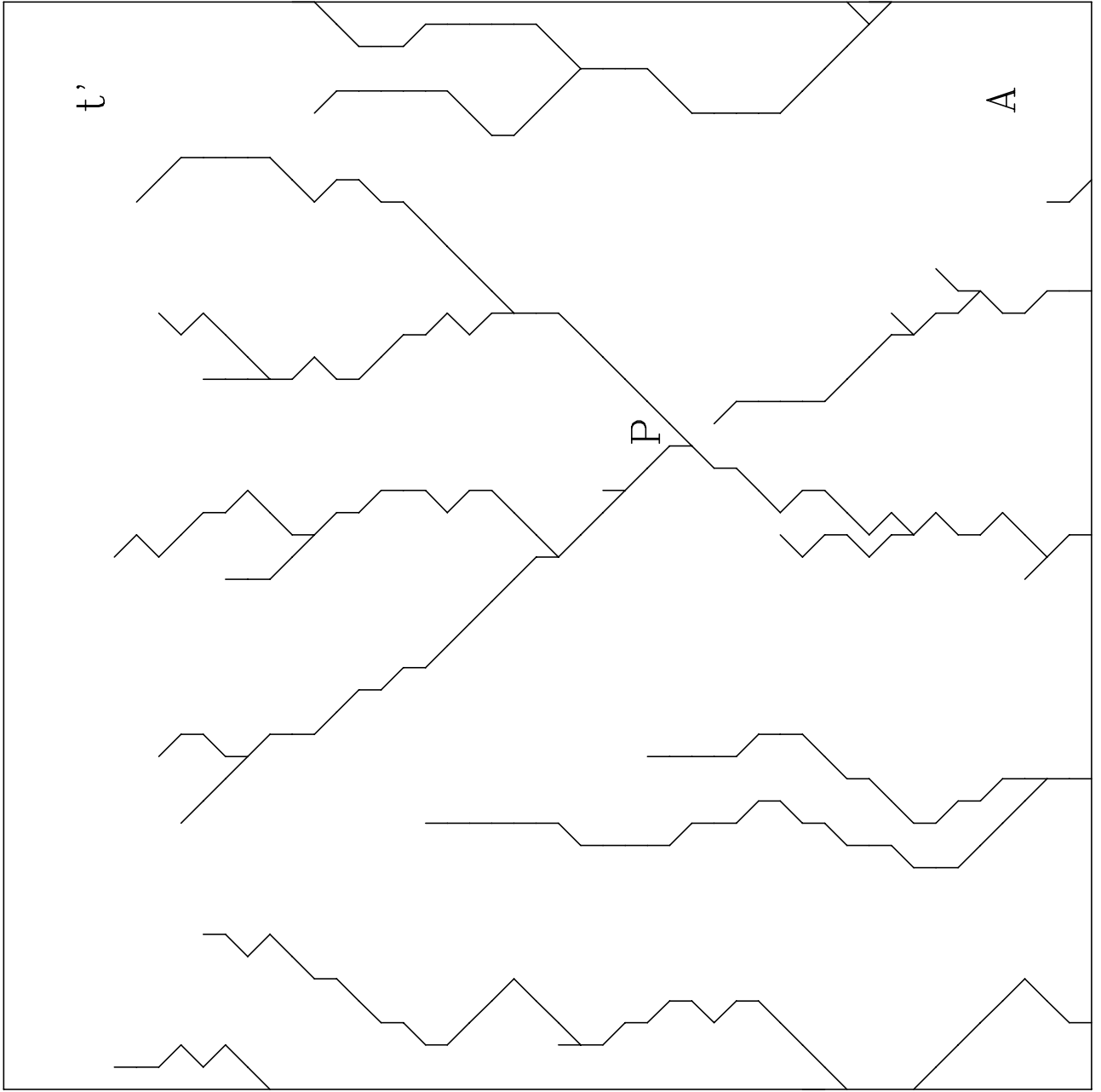
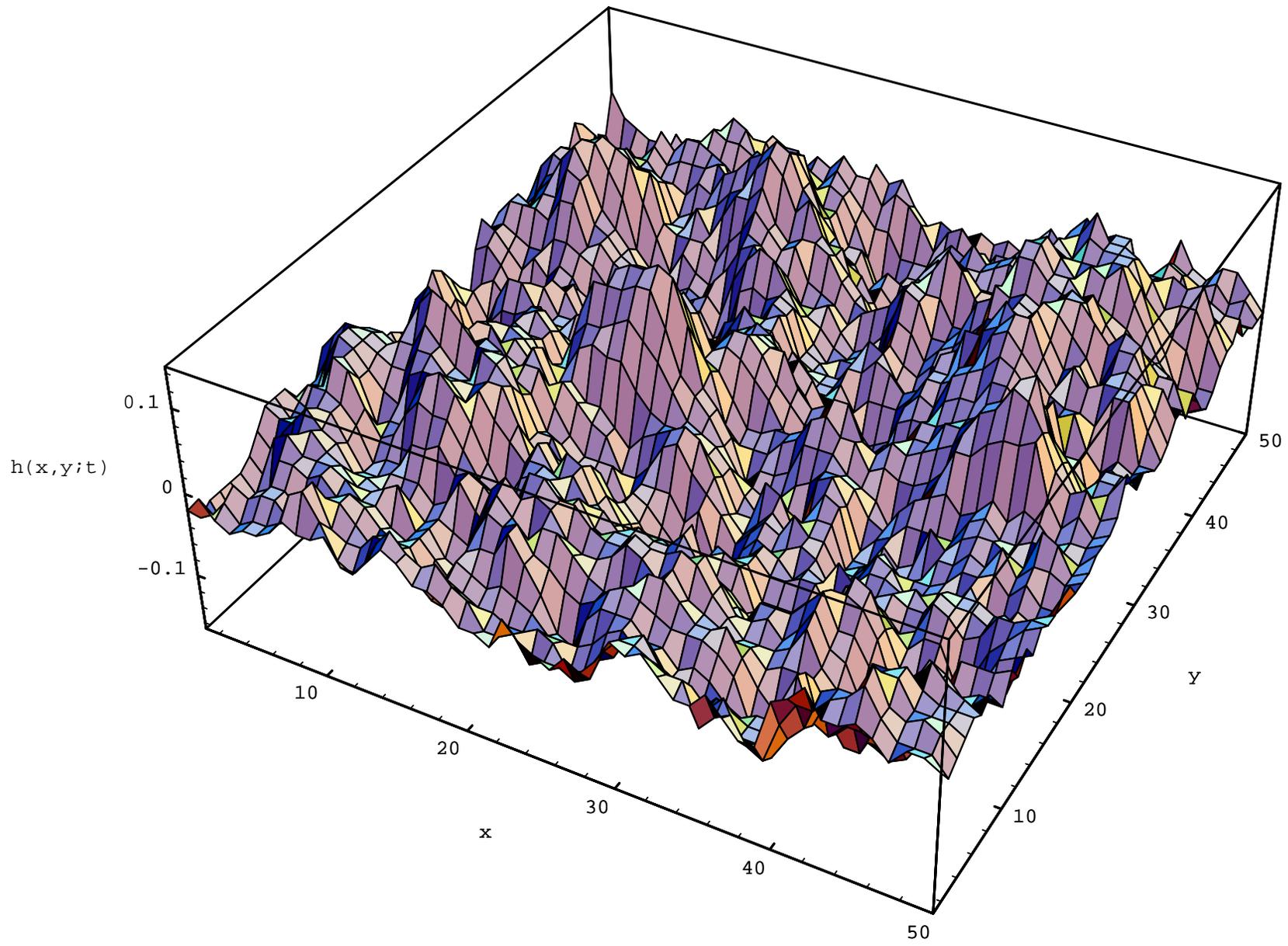
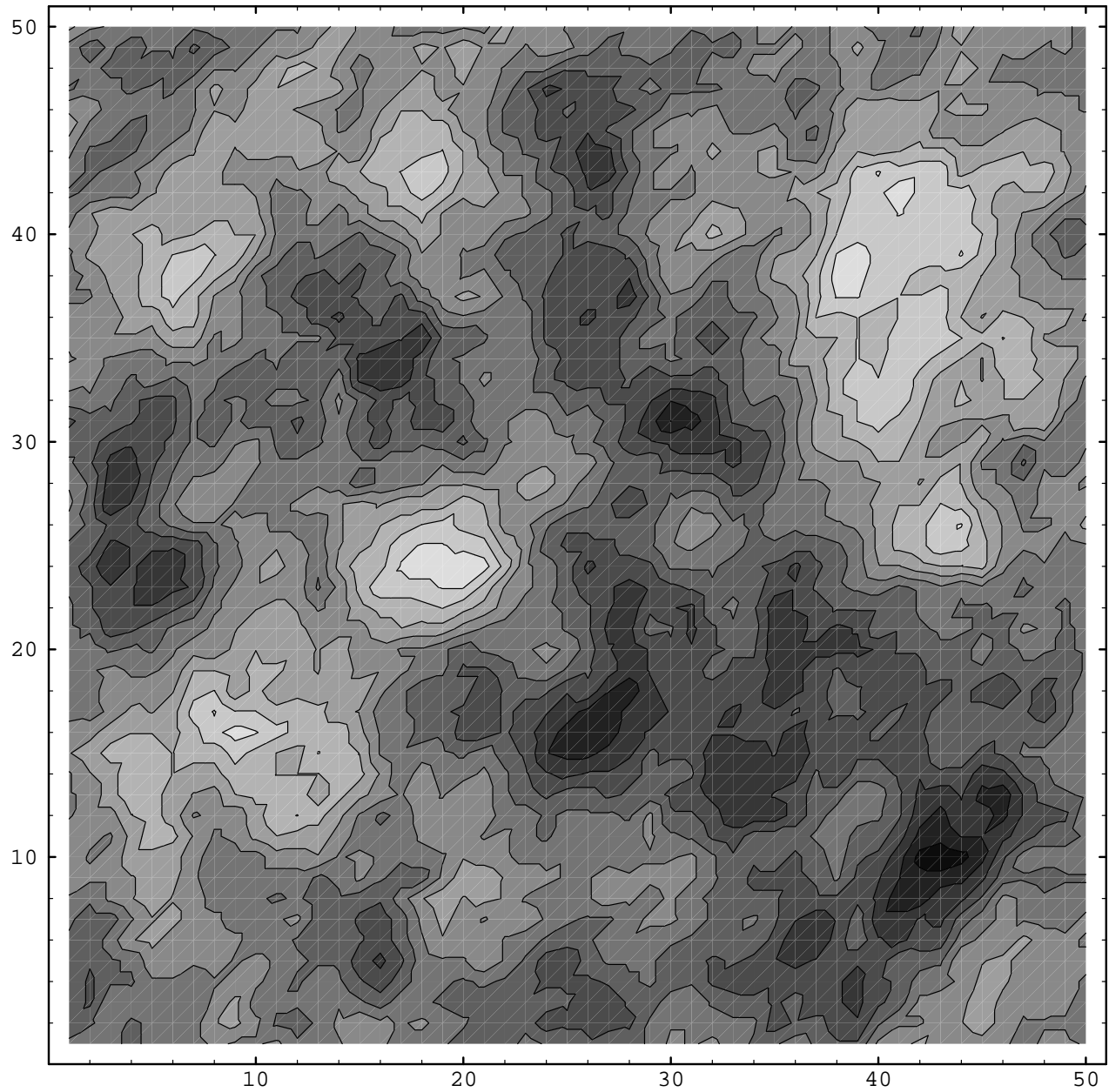
\end{document}